\documentclass{PoS}

\usepackage[]{latexsym}
\usepackage[]{amsfonts}
\usepackage[]{amsmath}
\usepackage[]{graphicx}
\usepackage[]{epsf}

\newcommand{\Tr}{\textrm{Tr}}

\newcommand{\ev}[1]{\langle #1 \rangle}

\newcommand{\dd}{\textrm{d}}
\newcommand{\LE}{\mathcal{L}_{\textrm{E}}}
\newcommand{\SE}{S_{\textrm{E}}}
\newcommand{\MSb}{\overline{\textrm{MS}}}

\newcommand{\mub}{\bar\mu}
\newcommand{\Nc}{N_{\textrm{c}}}
\newcommand{\Nf}{N_{\textrm{f}}}

\rightline{\sffamily CERN-PH-TH/2006-209}
\title{Quark number susceptibility of
high temperature 
QCD}

\ShortTitle{Quark number susceptibility at high temperature}

\author{\speaker{Ari Hietanen}\\
        Theoretical Physics Division, Department of Physical Sciences P.O.Box
        64 FI-00014 University of Helsinki, Finland and\\
	Helsinki Institute of Physics,	P.O.Box 64, FI-00014 University of Helsinki, Finland\\ 
        E-mail: \email{ari.hietanen@helsinki.fi}}

\author{Kari Rummukainen\\
        Department of Physics, University of Oulu P.O.Box 3000, FI-90014 Oulu,
        Finland and \\
        Department of Physics, Theory Division  CERN CH-1211 Geneva, Switzerland\\
        E-mail: \email{kari.rummukainen@oulu.fi}}

\abstract{We use three dimensional reduced effective field theory (EQCD) and lattice
calculations to determine the quark number susceptibility of QCD at high
temperature. We find our results to agree well  with known
perturbative expansion as well as with other lattice data.}  

\FullConference{XXIVrd International Symposium on Lattice Field Theory\\
                 25-30 July 2006\\
                 Tucson, Arizona, July 23-28, 2006}

\begin{document}

\section{Introduction}
A dimensionally reduced effective theory is, by now, a
well-established and powerful tool for studying high-temperature QCD
both perturbatively and non-perturbatively.  At high enough
temperatures the QCD coupling becomes small, and perturbative methods
can be safely applied to hard ($\sim T$) modes.  The fact that
high-temperature QCD can be reduced to 3d effective theory can be
understood by considering an Euclidean finite-$T$ propagator
\begin{equation}
  \frac{1}{{\boldmath p}^2 + \omega^2_n + m^2},
\end{equation}
where $ \omega_n^b=2n\pi T $ for bosons and $ \omega_n^f=(2n+1)\pi T $
for fermions.  Thus, only static $n=0$ bosonic modes remain ``light''
at high temperatures, and the heavy $n\neq 0$ bosonic and all
fermionic modes can be integrated perturbatively
\cite{ginsparg80,appelquist81,kajantie95,braaten95}.  The result is an
effective 3d theory of full QCD, electrostatic QCD (EQCD) \cite{klrs}.
An essential feature of the effective theory is that it fully includes
the perturbatively problematic soft ($\sim gT$) and ultrasoft ($\sim
g^2 T$) scales of the original theory; only perturbatively
well-controlled hard scales ($\sim T$) are integrated over.

EQCD offers a good starting point both for perturbative
calculations~\cite{gsixg,vuorinen03,gynther05} and non-perturbative
lattice simulations.  In the latter case, EQCD offers an interesting
alternative to standard high-temperature lattice simulations: above
all, the effective theory is purely bosonic and only 3-dimensional,
making it much cheaper to simulate.  The theory is
superrenormalizable, which renders the continuum limit particularly
transparent; it also enables simulations at arbitrarily large
temperatures.  On the other hand, the effective theory cannot be used
to study the QCD phase transition: at too low $T$ QCD becomes strongly
coupled and the perturbative derivation of EQCD fails.  Nevertheless,
the theory has been found to work well down to temperatures $T\sim
1.5-3\,T_c$, depending on observable
used.  Lattice simulations of EQCD have been used to calculate
QCD pressure at high T \cite{pressure}, spatial string tension
\cite{spatialstring}, and spatial screening lengths
\cite{hart00}.  Here we measure the quark number
susceptibility, and compare the result to both perturbative
and 4d lattice results.

\section{Susceptibility in electrostatic QCD}
EQCD is defined by the action
\begin{eqnarray}
       \SE & = & \int \dd^3x\LE \nonumber \\
      \LE & = & \frac{1}{2}\Tr[F_{ij}^2]+\Tr[D_i,A_0]^2+m_3^2\Tr[A_0^2]+
      i\gamma_3 \Tr[A_0^3] + \lambda_3(\Tr[A_0^2])^2,
      \label{action}
\end{eqnarray}
where $F_{ij}=\partial_i A_j - \partial_j A_i + ig_3[A_i,A_j]$ and
$D_i=\partial_i+ig_3A_i$. $F_{ij}$, $A_i$ and $A_0$ are traceless
$3\times 3$ Hermitean matrices ($A_0=A_0^aT_a$, etc).  Coupling and
mass parameters $g_3$, $m_3$, $\gamma_3$ and $\lambda_3$ are defined
by the physical 4d temperature, renormalization scale
$\Lambda_{\MSb}$, chemical potential $\mu$ and 
the number of massless fermions.  It is convenient to use
the dimensionless ratios
\begin{equation}
  y  =  \frac{m_3^2}{g_3^4}, \;\; x = \frac{\lambda_3}{g_3^2},
  \;\; z=\frac{\gamma_3}{g_3^3},
\end{equation}
which determine the physical properties of EQCD.  The 
$\mu$-dependence of the parameters is, at 1-loop level,
\begin{equation}
         y = y_{\mu=0} \left(1+\sum_f\mub_f^2\frac{3}{2\Nc+\Nf}\right),
  ~~~~~~ z = \sum_f \frac{\mub_f}{3\pi},
  ~~~~~~ x = x_{\mu=0}\,,
\end{equation}
where $\mub = \mu/(\pi T)$ and the $\mu=0$ expressions can 
be found in ref.~\cite{klrs}.
The two loop corrections have been calculated in ref.~\cite{hart00},
but the effects remain in practice negligible. 

The quantity we are interested in is the quark number susceptibility,
which we calculate over one flavor $u$ only, can be defined in EQCD as 
\begin{equation}
  \chi_3  =   \frac{1}{V}\frac{\partial^2}{\partial \mu_u^2}\ln{\cal Z} 
   =   \frac{1}{V}\frac{\partial^2}{\partial \mu_u
    ^2}\ln\int \mathcal{D}A_kA_0\exp\left(-\SE\right)
\end{equation}
Substituting $\SE$ from (\ref{action}) we arrive at equation
\begin{eqnarray}
  \chi_3& = & -\frac{6}{2N_c+N_f}\,y_{\mu=0}\, \ev{\Tr A_0^2} +
  V\frac{N_f^2}{9\pi^2} \ev{(\Tr A_0^3)^2} \nonumber \\
  & & +V \frac{9N_f^2}{(2N_c+N_f)^2}\mub^2\,y_{\mu=0}^2\,
  \left(\ev{(\Tr A_0^2)^2} - \ev{\Tr A_0^2}^2\right) 
\end{eqnarray}
Thus, the quark number susceptibility is obtained by measuring the
condensates $\ev{Tr A_0^2}$, $\ev{(\Tr A_0^2)^2}$ and 
$\ev{(\Tr A_0^3)^2}$ on the lattice.  Due to the
superrenormalizable nature of the theory,  measurements can
be rigorously converted to $\MSb$ scheme in the lattice continuum limit;
because $\MSb$ was used in in the perturbative matching to 4d QCD,
this also allows us to compare to 4d results.
For example, the continuum limit for the $ \ev{[Tr A_0^3]^2}$ is
\begin{eqnarray}
  V\ev{[\Tr A_0^3/g_3^3]^2}_{\MSb} & = & \lim_{\beta \rightarrow
  \infty} \left\{V\ev{[\Tr A_0^3/g_3^3]^2]}_a-\frac{5}{16\pi^2}\left[\ln\beta+0.08848010\right]\right\},
\end{eqnarray}
where $\beta\equiv \frac{2\Nc}{ag_3^2}$ is the lattice coupling
constant. Continuum limits for $\ev{Tr A_0^2}$ and $\ev{(\Tr
A_0^2)^2}$ are given in \cite{klrs}.  The relation between $\chi_3$
and the true 4d susceptibility is
\begin{equation}
  \chi = \frac{g_3^6}{T^3}\chi_3 + \frac{\partial^2}{\partial
  \mu_u^2}\Delta p,
  \label{eqsusc}
\end{equation}
where $\Delta p=p_\textrm{QCD}-p_\textrm{3d}$ is the perturbative
3d$\rightarrow$4d matching coefficient, and can be found in
\cite{vuorinen03}. 

\section{Lattice measurements}

Lattice simulations were carried out for $\Nf=2$. We used large lattice
sizes $V=140^3,200^3$ and values of $\beta=$32, 40, 67, 80, 120.  
The volumes are large enough so that the remaining (exponential)
finite volume effects are safely below our statistical accuracy; for
a detailed analysis in a related theory see \cite{plaquette}.

For each value of $\beta$ we did simulations with values of
$y=6.62,\;5.31,\;4.00,\;3,09,\;2.02,\;1.18$, $0.71,$ and $0.45$. The 
simulations have been carried out for $\mu=0$, and to obtain
the susceptibility, one needs measurements of 
$\ev{\Tr A_0^2}$ and $\ev{(\Tr A_0^3)^2}$. 

\begin{figure}
  \begin{center}
    \includegraphics*[width=\textwidth]{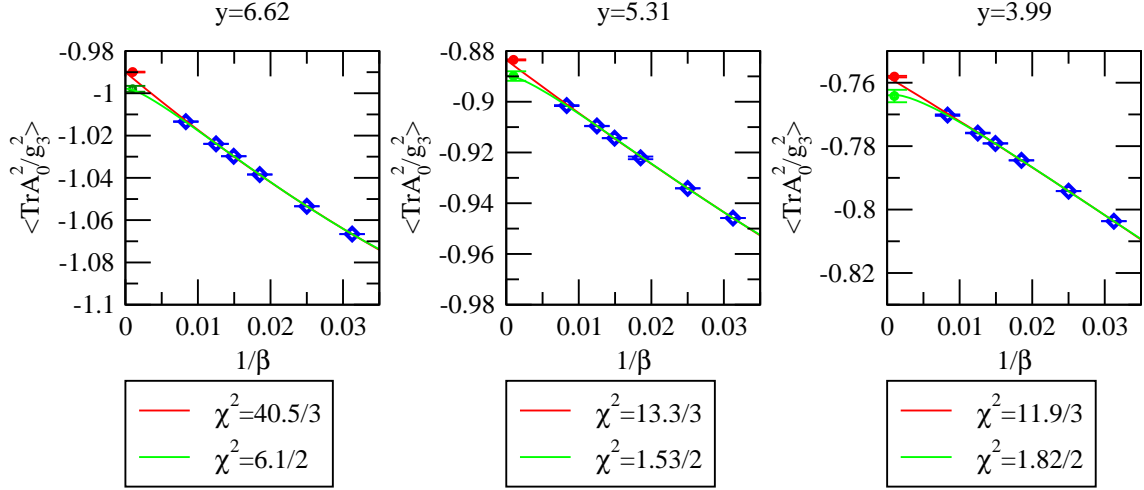}
    \caption{Difference of continuum extrapolations with and without
      the log-term. The solid line is a fit
      $c_1+\frac{c_2}{\beta}+\frac{c_3}{\beta^2}$ and the dashed line
      $c_1+\frac{c_2}{\beta}+{\frac{c_2'}{\beta}\log(\beta)}+\frac{c_3}{\beta^2}$.}
    \label{contexpa2}
  \end{center}
\end{figure}

\begin{figure}
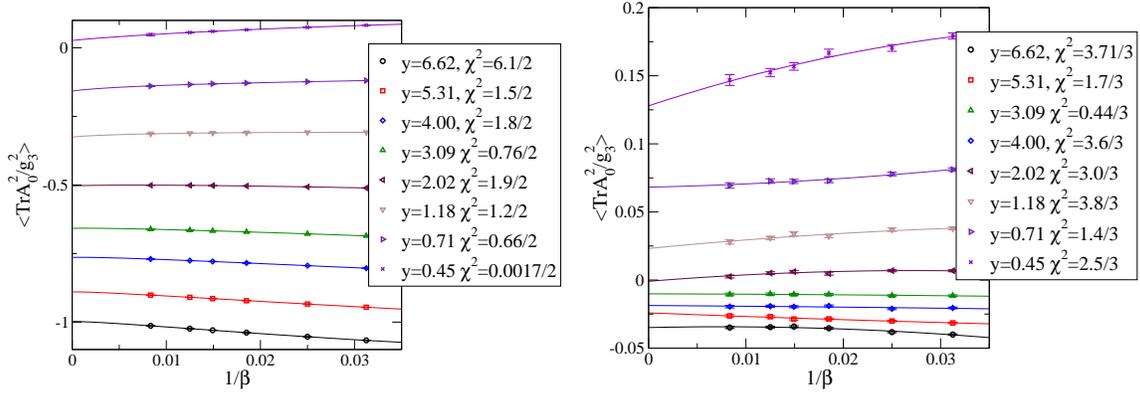

    \begin{center}
      \includegraphics*[width=210pt]{a2cl.eps} ~
      \includegraphics*[width=210pt]{a6cl.eps}
    \end{center}
    \caption{Continuum limits of $\ev{\Tr A_0^2}$ and
      $\ev{(\Tr A_0^3)^2}$ at different values of $y$.}
    \label{contexpa1}
\end{figure}

The standard procedure of doing continuum extrapolation is to fit a
polynomial to the divergence subtracted lattice data. However, in EQCD
the divergent lattice contributions contain also terms of the form
$\beta \log(\beta)$ and such terms could also arise in terms $\sim a \sim
1/\beta$.  Therefore, we do continuum extrapolation in two ways;
we use a second order polynomial
\begin{equation} 
  c_1+\frac{c_2}{\beta}+\frac{c_3}{\beta^2},
  \label{polyextra}
\end{equation}
and a fitting function of the form
\begin{equation} 
  c_1+\frac{c_2}{\beta}+{\frac{c_2'}{\beta}\log(\beta)}+\frac{c_3}{\beta^2}.
\label{logextra}
\end{equation}
The Fig.~\ref{contexpa2} demonstrates the differences between
different continuum extrapolations for $\ev{\Tr A_0^2}$. The
extrapolations done using (\ref{logextra}) have excellent $\chi^2/$dof,
and we shall use this form henceforth.
The full continuum extrapolations are shown in Fig.~\ref{contexpa1}.
We note that the detailed form of the fitting function is significant
for our results; thus, knowledge of the true $1/\beta$ coefficient is
highly desirable.  There is an ongoing calculation of this term using
stochastic perturbation theory \cite{torrero06}, which will hopefully
confirm our result.

The contributions of $\ev{(\Tr A_0^3)^2}$ to susceptibility are
numerically much smaller than that of $\ev{\Tr A_0^2}$. The
measurements of $\ev{(\Tr A_0^3)^2}$ are not accurate enough to
distinguish between extrapolations (\ref{polyextra}) and
(\ref{logextra}), and the choice is not significant for the final
results.  Hence we use the second order function in
Fig.~\ref{contexpa1}. 

\begin{figure}
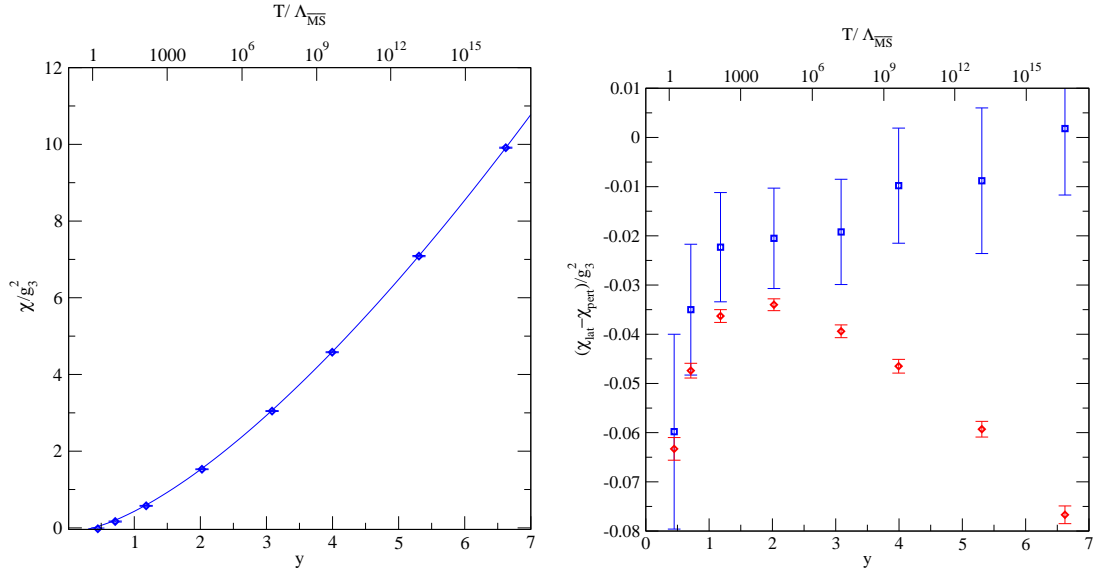

  \begin{center}
    \includegraphics*[width=200pt]{latsuskis.eps} ~~
    \includegraphics*[width=200pt]{suskisdelta.eps}
  \end{center}
  \caption{{\em Left:} the susceptibility as a function of $y$ and
    $T/\Lambda_{\MSb}$. The solid line is the perturbative prediction.
    {\em Right:} the difference between the perturbation theory and lattice
    results. Points denoted by squares are obtained from
    logarithmic continuum
    extrapolation and diamonds from polynomial extrapolation; the
    former one has expected behavior as $y\rightarrow \infty$. }
  \label{latsuskis}
\end{figure}

The continuum extrapolated results are shown in
Fig.~\ref{latsuskis}, as a function of the parameter 
$y(T)$.    The result agrees very well with 
the perturbative susceptibility derived in ref.~\cite{vuorinen02}. 
The perturbative result has the form:
\begin{equation}
  \chi_{\textrm{pert}}=a_1y^{3/2}+a_2y+a_3y^{1/2}+a_4
\end{equation}
where $a_i$ are functions of $x$ and $z$.  On the right panel
we show the difference between the perturbative result and the
lattice measurement.  Because the lattice measurement (in the continuum
limit) fully includes perturbative contributions, we expect the
difference 
to behave as $O(y^{-1/2})$.  This is the case when continuum
extrapolation is performed using Eq.~(\ref{logextra}); the extrapolation
(\ref{polyextra}) diverges from perturbative result as $y\rightarrow 
\infty$, which is inconsistent.  This also motivates the use
of (\ref{logextra}) as continuum extrapolation.

\begin{figure}
  \begin{center}
    \includegraphics*[width=200pt]{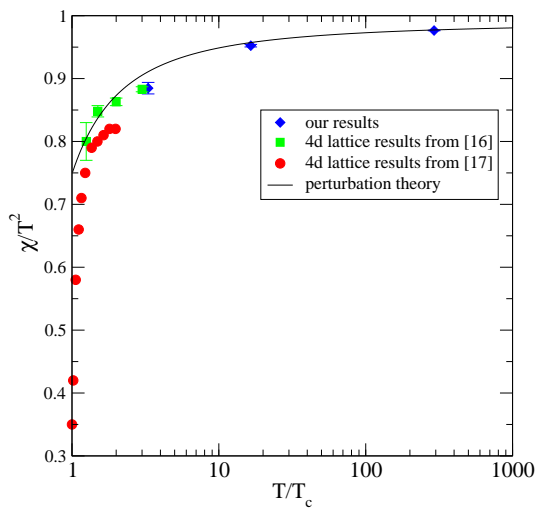}
  \end{center}
  \caption{The quark number susceptibility in 4d units.
    Our results are shown with diamonds, only 3 points fit in the 
    temperature range shown.  The results agree very well with
    4d lattice simulation results \cite{gavai,karsch}.}
  \label{fig4d}
\end{figure}

In Fig.~\ref{fig4d} we show the susceptibility in 4d units,
transformed using Eq.~(\ref{eqsusc}), and compare the result to 4d
$\Nf=2$ staggered fermion lattice simulations \cite{gavai,karsch}.  Here we use the ratio
$T_c/\Lambda_{\MSb} =0.49$ \cite{gupta}.  We can observe that our results agree
very well with 4d simulations and with perturbation theory.  Indeed,
the difference between EQCD measurements and perturbative results are
all but invisible on the scale of the plot, possibly excluding the
lowest temperature data.  


\section{Conclusions and Outlook}

We have measured the quark number susceptibility in dimensionally reduced
effective theory of high-$T$ QCD.  The results show 
a good agreement with the perturbation theory and 4d lattice simulations,
down to temperatures $\sim 3 T_c$.  
Currently we are expanding our simulations for finite chemical 
potential.  In order to avoid the sign problem at finite $\mu$
we do the simulations using imaginary chemical potential and
then continue analytically to real values. 

\acknowledgments
This work was partly supported by the Magnus Ehrnrooth Foundation, a
Marie Curie Host Fellowship for Early Stage Researchers Training, and
Academy of Finland, contract number 104382. Simulations were carried
out at Finnish  Center for Scientific Computing (CSC).


\end{document}